\documentclass[preprint,aps,showpacs]{revtex4}
\usepackage{graphicx}
\usepackage{dcolumn}
\usepackage{bm}
\preprint{}
\def\para{\par\noindent}
\def\sqr#1#2{{\vcenter{\vbox{\hrule height.#2pt
        \hbox{\vrule width.#2pt height#1pt \kern#1pt
          \vrule width.#2pt}
        \hrule height.#2pt}}}}

\newcount\notenumber

\def\note{\advance\notenumber by 1
\footnote{$^{\the\notenumber}$}} \baselineskip 20pt
\begin{document}
\title{Persistence and the Random Bond Ising Model in Two Dimensions}
 \author{S.~Jain}
 \affiliation{Information Engineering \\
The Neural Computing Research Group \\ School of Engineering and Applied Science \\
Aston University \\ Birmingham B4 7ET \\ U.K.}
 \author{H.~Flynn}
 \affiliation{School of Mathematics and Computing\\
 University of Derby\\
 Kedleston Road\\
 Derby DE22 1GB\\ U.K.}

\begin{abstract}

We study the zero-temperature persistence phenomenon in the random
bond $\pm J$ Ising model on a square lattice via extensive numerical
simulations. We find strong evidence for \lq blocking\rq\ regardless
of the amount disorder present in the system. The fraction of spins
which {\it never} flips displays interesting non-monotonic,
double-humped behaviour as the concentration of ferromagnetic bonds
$p$ is varied from zero to one. The peak is identified with the onset of the zero-temperature
spin glass transition in the model. The residual persistence is found to
decay algebraically and the persistence exponent $\theta (p)\approx
0.9$ over the range $0.1\le p\le 0.9$. Our results are completely consistent
with the result of Gandolfi, Newman and Stein for infinite systems that this model
has \lq mixed\rq\ behaviour, namely positive fractions of spins that
flip finitely and infinitely often, respectively. [Gandolfi, Newman
and Stein, Commun. Math. Phys. {\bf 214} 373, (2000).]
\end{abstract}

\maketitle
\section{Introduction}
\para In recent years there has been considerable interest in the
\lq persistence\rq\ problem [1-11] and it has been studied theoretically in a wide
range of systems. Generically, this problem is
concerned
 with the fraction of space
which persists in its initial ($t=0$) state up to some later time.
\para Thus, when studying the non-equilibrium dynamical behaviour of spin systems
 at zero-temperature we
are interested in the fraction of spins, $P(t)$, that persists for $t>0$ in the
same state
as at $t=0$.
\para It has now been established for quite sometime that for the pure
ferromagnetic two-dimensional Ising model, $P(t)$ decays algebraically [1-4]
$$P(t)\sim t^{-\theta }, \eqno(1)$$
 where $\theta = 0.209\pm 0.002$ [5]. Similar algebraic decay has been found in
numerous other systems displaying persistence [10, 11].
\para However, computer simulations of the Ising model in high dimensions [3], $d > 4$,
 the $q-$state Potts [12] ($q > 4$) have suggested the presence of a
 non-vanishing persistence probability as $t\to\infty$; this feature is
 sometimes referred to as \lq blocking\rq\ and has also been found to
 be present in some models containing disorder [5-6, 13-15].
 Clearly, if $P(\infty) > 0 $, the problem can be reformulated by
restricting attention only to those spins that eventually do flip.
Therefore, we can study
 the behaviour of the residual persistence
$$r(t)=P(t)-P(\infty).\eqno(2)$$
Most of the initial effort was restricted to studying pure systems
and, it\rq\ s only fairly recently that the persistence behaviour of
systems containing disorder has been studied [5-6, 13-15]. Very
recently [16], the local persistence exponent for the axial
next-nearest neighbour Ising model has been estimated to be $\theta
=0.69\pm 0.01$; a value considerably different to that found for the
ferromagnetic Ising model.

Numerical simulations of the bond diluted Ising model [5,6] indicate
that the long time behaviour of the system depends on the amount of
disorder present. For the weakly diluted system [5], there is
evidence of non-algebraic decay prior to \lq blocking\rq . For the
strongly diluted model, on the other hand, the residual persistence
probability decreases exponentially for large times [6].

Although the presence of a \lq blocked\rq\ state has also been
suggested [13-15] for the random bond Ising model in $2d$, the long
time behaviour of the residual persistence has not been investigated
to-date and is still an open question.

\para In this work we attempt to fill the gap by presenting new results
of extensive computer
simulations of the $2d$ random bond Ising model on a square lattice
for a wide range of bond concentrations. Our main objective is to investigate the
persistence behaviour as a function of the ferromagentic bond concentration.
  As we shall see, we find
strong evidence for \lq blocking\rq\ regardless of the amount of
disorder present in the system. Furthermore, unlike the bond-diluted
case [5,6], here the qualitative behaviour of the model does not
appear to depend on the concentration of the disorder.

\para In the next section we introduce the model and give brief details
about the method used to perform the simulations. In section III we
discuss the results and finish with some concluding remarks.
\section{The Model}
\para The Hamiltonian for our model is given by
$${\it H} = -\sum_{<ij>}  J_{ij}S_i S_j \eqno(3)$$
where $S_i=\pm 1$ are Ising spins situated on every site of a
 square lattice with periodic boundary conditions and the
quenched ferromagnetic exchange interactions are selected from
 a binary distribution given by
$$P(J_{ij}) = (1-p)\delta(J_{ij}+ J) + p \delta(J_{ij}-J)\eqno(4)$$
where $p$ is the concentration of ferromagnetic bonds and we set $J=1$;
 the summation in Eqn. (3)
runs
 over all nearest-neighbour pairs only. Note that for $p=1/2$ and $p=1$ we obtain the
Ising spin glass and the pure ferromagnetic Ising models,
respectively.

Initial runs for a range of ferromagnetic bond concentrations were
performed for lattices of linear dimensions ranging from $L=250$ to
$L=1000$. No appreciable finite-size effects were evident for the range of values considered.
As a
result, the data presented in this work were obtained for a lattice
with dimensions
 $500\times 500\ (=N)$.
\para Each simulation run begins at $t=0$ with a random
starting configuration
of the spins and
then we update the lattice via single spin flip zero-temperature
 Glauber dynamics [5].
 The updating rule we use is: always flip if the energy
 change is negative, never flip if the energy change is positive and flip at
random if the energy change is zero.

\para  The number, $n(t)$,
of spins which have never flipped until time $t$ is then counted.

\para The persistence
probability is defined by [1]
$$P(t)=[<n(t)>]/N\eqno(5)$$
where $<\dots>$ indicates an average over different initial
conditions and $[\dots]$ denotes an average over samples. Averages
over at least 100 different initial conditions and samples were
performed for each run undertaken.
\section{Results}
\para We now discuss our results. The behaviour of the persistence
probability is displayed in the log-log plot shown in Figure 1
($0.1\le p\le 0.5$). The problem is symmetric about the spin-glass ($p=0.5$)
case and, as a check on the numerics, we confirmed that similar plots were obtained
for ($0.5\le p\le 0.9$). Note that the
error-bars are smaller than the size of the data points. It\rq s
clear from the figure that $P(t)$ is finite in the long time limit.
Hence, the system is \lq blocked\rq . The \lq blocking\rq\
probability depends on $p$, the concentration of ferromagnetic bonds
present. However, it would appear that \lq blocking\rq\ occurs for
all of the values of $p$ considered.

\begin{figure}
\includegraphics{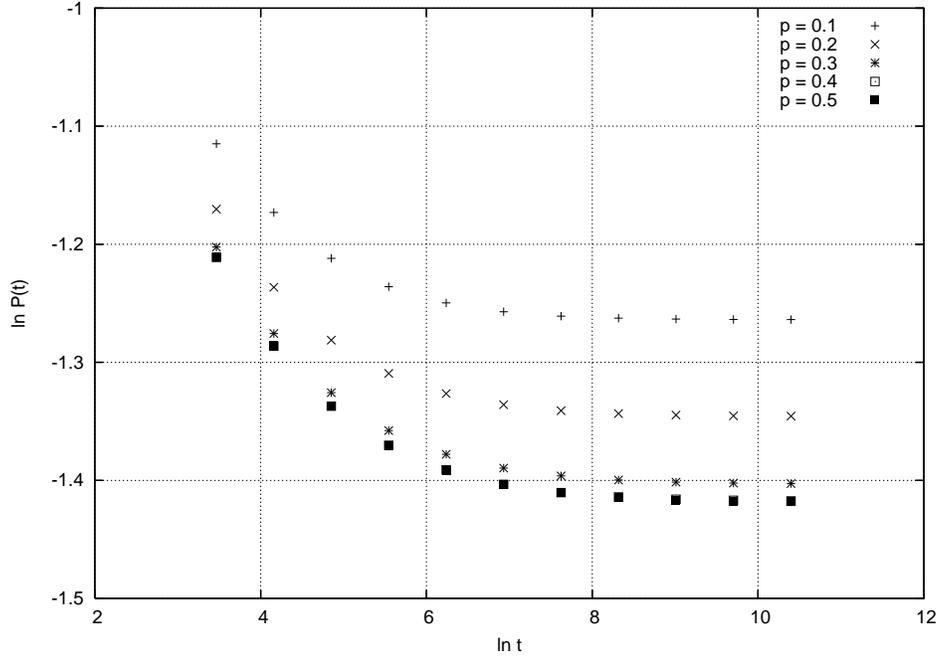}
\caption{A log-log plot of the persistence against time for a range
of bond concentrations, $0.1\le p\le 0.5$. Note that the data for $p
= 0.5$ are superimposed over those for $p = 0.4$.} \label{fig1}
\end{figure}

To explore the \lq blocking\rq\ feature further, we plot in Figure 2
the data over a narrow range very close to the pure case, namely
$0.95\le p\le 0.999$. For reference purposes, the straight line in
Figure 2 has a slope of $-0.21$ and corresponds to the well
established persistence exponent for the pure case $p = 1.0$. It\rq
s clear from Figure 2 that we have deviations from the pure case
even when $p = 0.999$. For all values of ferromagnetic bond
concentrations $p\ne 1$ we have a finite fraction of spins which
{\it never} flips. Furthermore, the blocking probability,
$P(\infty)$, also appears to be highly sensitive to the value of
$p$.
\begin{figure}
\includegraphics{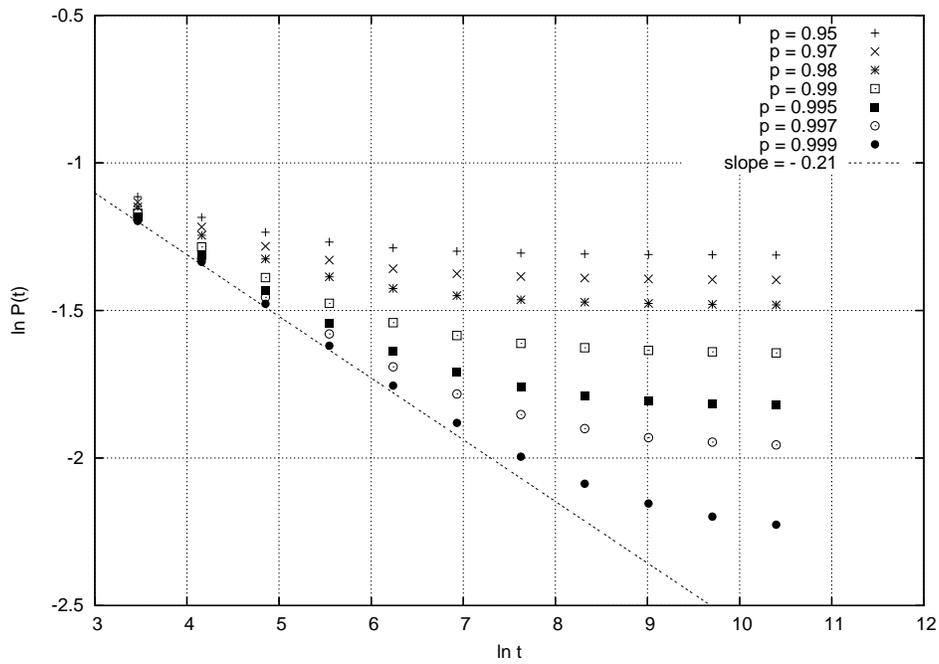}
\caption{A plot of $\ln P(t)$ against $\ln t$ for $0.95\le p\le
0.999$. The straight line, corresponding to the behaviour for the
pure ($p = 1.0$) case, has gradient $-0.21$.} \label{fig2}
\end{figure}

In order to examine the behaviour of the \lq blocking\rq\
probability, we plot in Figure 3 the values extracted for
$P(\infty)$ from Figures 1 and 2 against the bond concentration. Note
that Figure 3 shows the values of $P(\infty)$ for a wide range of
$p: 0\le p\le 1$, including some values which have not been
displayed in the earlier figures for clarity. Once again, the symmetry of the plot
about $p = 0.5$ acts as a consistency check on the numerics. The plot itself appears to
have an interesting non-monotonic, double-humped feature. In our model the average fraction of
frustrated plaquettes, $Plaq_f$, is given by [17]
$$Plaq_f = 4p(1-p)(p^2+(1-p)^2)\eqno(6)$$
and there is a zero-temperature spin glass transition at $p_c\approx
0.11$ [18]. We see from Figure 3 that the peak in the blocking
probability coincides with this value of $p_c$. Furthermore, $Plaq_f
(p_c\approx 0.11)\approx 0.31 < 1/2$, the maximum value of $Plaq_f$.
\begin{figure}
\includegraphics{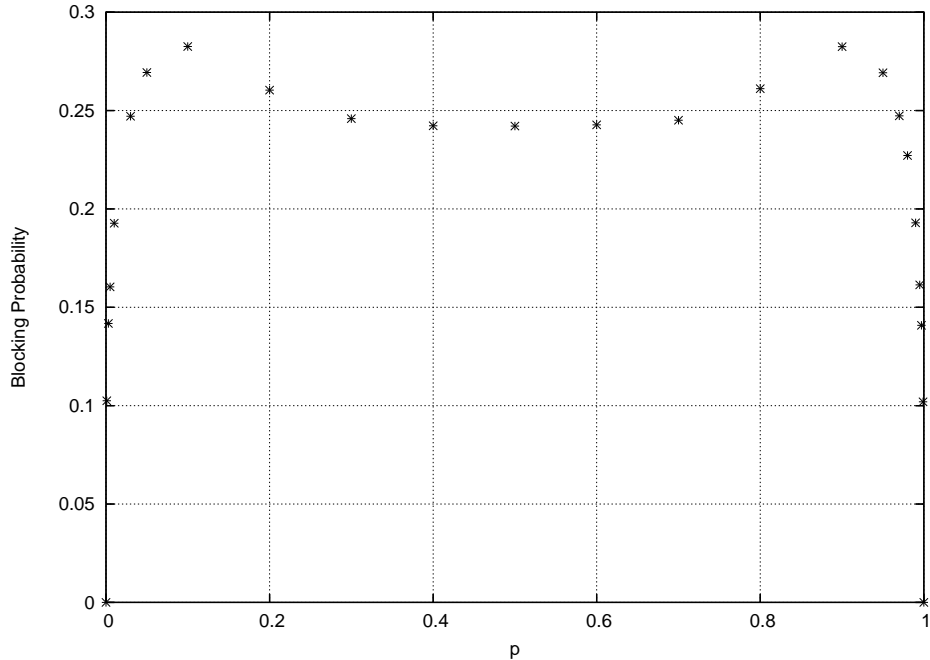}
\caption{A plot of the \lq Blocking Probability\rq\ , $P(\infty)$,
against the bond concentration, $p$.} \label{fig3}
\end{figure}

As explained earlier, for a blocked system, we can study the
residual persistence $r(t) = P(t) - P(\infty)$. After having
extracted the blocking probability for each $p$, we calculate
$r(t)$. However, there is an error involved in estimating
$P(\infty)$. As a consequence, the error in $r(t)$ is much greater
than that in the original persistence probability. In Figure 4 we
show log-log plots of $r(t)$ against $t$ for three selected values
of $p = 0.1, 0.5$ and $0.9$. In each case, we see that the decay of
the residual persistence is algebraic over the time-interval
concerned. However, because of the uncertainty in the blocking
probability, there are not inconsiderable error bars attached to the
resulting (residual) persistence exponents.
\begin{figure}
\includegraphics{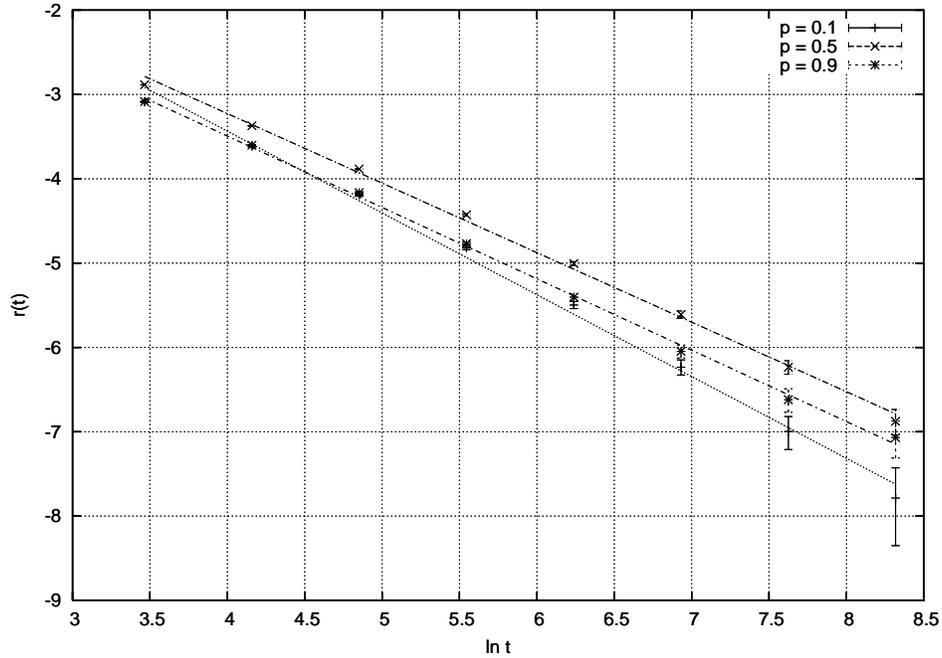}
\caption{Here we show a log-log plot of the residual persistence
against time for selected bond concentrations. The straight lines
are guides to the eye and indicate that $\theta = 0.97(8), 0.83(3),
0.85(4)$ for $p = 0.1, 0.5, 0.9$, respectively.} \label{fig4}
\end{figure}

Our estimates for the persistence exponents, $\theta (p)$, are
plotted against the bond concentrations in Figure 5. It would appear
that $0.8\le \theta (p)\le 1.1$ when $0.1\le p\le 0.9$. For
reference, the exponent for the pure case is indicated by the arrow.
Note that although our data are not influenced by finite-size effects, it\rq s
nevertheless a non-trivial matter to extract the residual persistence exponent
for $0<p<0.1$ because of the sensitivity to the estimate of $P(\infty)$. As can be
deduced from Figure 2, the closer we are to the pure case, the more difficult it is
to estimate the blocking probability.
\begin{figure}
\includegraphics{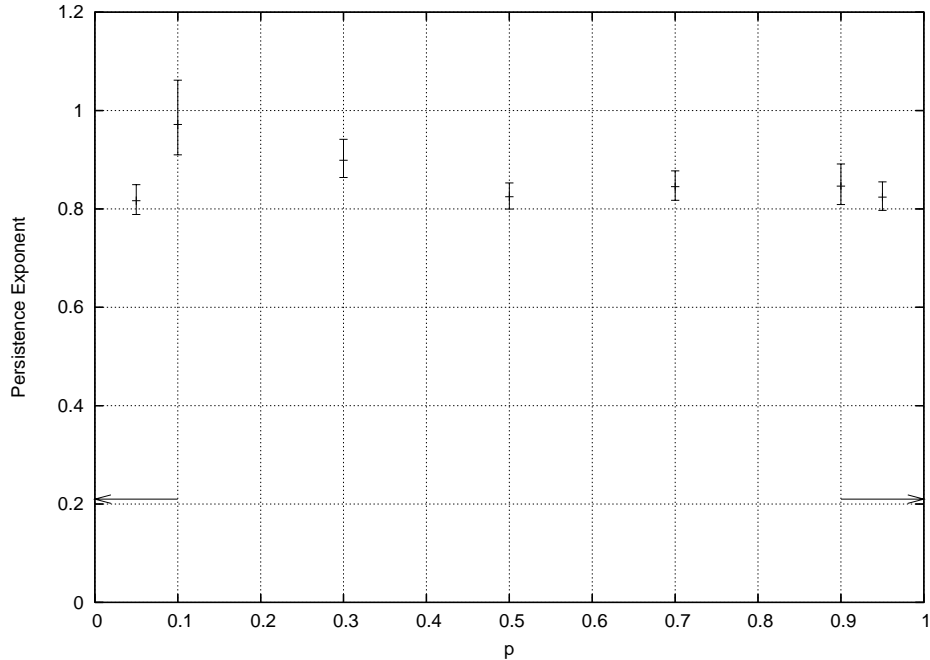}
\caption{A plot of $\theta (p)$ against $p$. For reference, the
arrows indicate where the pure, ferromagnetic ($p = 1.0$) and
anti-ferromagnetic ($p = 0$), values appear on the plot.}
\label{fig5}
\end{figure}
\section{Conclusion}
\para To conclude, we have presented new data for the random bond Ising
models on a square lattice. Our results confirm the existence of \lq
blocking\rq\ in the system regardless of the amount of disorder
present. The results are consistent with the presence of positive
fractions of spins which flip finitely and infinitely often,
respectively. We have also investigated the \lq blocking\rq\
probability and find interesting non-monotonic behaviour as a
function of the ferromagnetic bond concentration. The persistence
exponent has been extracted and found to be $\approx 0.9$,
independent of the bond concentration over the range $0.1\le p\le
0.9$. Although we know that $\theta (p=1.0) =0.209\pm 0.002$ [5],
an accurate extraction of the residual persistence exponent for $p$ close to $1$
is highly sensitive to the (assumed) value for $P(\infty)$.
As a consequence, the complete nature of $\theta (p)$ over the entire range of bond
concentrations remains to be established.
\section {Acknowledgement}
\para HF would like to thank the University of Derby for a Research
Studentship while this work was carried out.
\section*{References}
\begin{description}
\item {[1]} B. Derrida, A. J. Bray and C. Godreche, J.Phys. A {\bf 27},
 L357 (1994).
\item {[2]} A.J. Bray, B. Derrida and C. Godreche, Europhys. Lett. {\bf 27},
 177 (1994).
\item {[3]} D. Stauffer J.Phys.A {\bf 27}, 5029 (1994).
\item {[4]} B. Derrida, V. Hakim and V. Pasquier, Phys. Rev. Lett. {\bf 75},
 751 (1995); J. Stat. Phys. {\bf 85}, 763 (1996).
\item {[5]} S. Jain, Phys. Rev. E{\bf 59}, R2493 (1999).
\item {[6]} S. Jain, Phys. Rev. E{\bf 60}, R2445 (1999).
\item {[7]} S.N. Majumdar, C. Sire, A.J. Bray and S.J. Cornell, Phys. Rev. Lett.
 {\bf 77}, 2867 (1996).
\item {[8]} B. Derrida, V. Hakim and R. Zeitak, Phys. Rev. Lett. {\bf 77}
 2971 (1996).
\item {[9]} S.N. Majumdar and A.J. Bray, Phys. Rev. Lett. {\bf 81} 2626 (1998).
\item {[10]} S.N. Majumdar, {\it Curr. Sci.} {\bf 77} 370 (1999).
\item {[11]} P. Ray, Phase Transitions {\bf 77} (Nos. 5-7), 563 (2004).
\item {[12]} B. Derrida, P.M.C. de Oliveira, D. Stauffer, Physica A{\bf 224}, 604 (1996).
\item {[13]} C.M. Newman and D.L. Stein, Phys. Rev. Lett. {\bf 82},
3944 (1999).
\item {[14]} C.M. Newman and D.L. Stein, Physica A{\bf 279}, 159
(2000).
\item {[15]} A. Gandolfi, C.M. Newman and D.L. Stein, Communication
Math. Phys. {\bf 214}, 373 (2000).
\item {[16]} P. Sen and S Dasgupta, J. Phys. A {\bf 37}, 11949
(2004).
\item {[17]} S. Kirkpatrick, Phys. Rev. B{\bf 16}, 4630 (1977).
\item {[18]} N. Kawashima and H. Rieger, Europhys. Lett. {\bf 39}, 85 (1997).
\end{description}
\end{document}